\documentclass[twocolumn,showpacs,preprintnumbers,amsmath,amssymb,prl]{revtex4-1}

\usepackage{graphicx}% Include figure files
\usepackage{dcolumn}% Align table columns on decimal point
\usepackage{bm}% bold math

\begin{document}

%\preprint{APS/123-QED}

\title{Polarity patterns of stress fibers}

\author{N.~Yoshinaga}
\altaffiliation{Present address: 
Fukui Institute for Fundamental Chemistry, Kyoto
University, Kyoto 606-8103, Japan}
\author{J.-F.~Joanny}
\author{J.~Prost}
\altaffiliation{E.S.P.C.I., 10 rue Vauquelin, 75005 Paris, France}
\author{P.~Marcq}
\email{philippe.marcq@curie.fr}
\affiliation{
Physico-Chimie Curie, 
Institut Curie, CNRS, Universit\'e Pierre et Marie Curie,
26 rue d’Ulm, F-75248 Paris Cedex 05 France
}

\date{November 13, 2010}
%\date{\today}

\begin{abstract}
Stress fibers are contractile actomyosin bundles commonly observed in 
the cytoskeleton of metazoan cells. The spatial profile of the polarity 
of actin filaments inside contractile actomyosin bundles
is either monotonic 
(graded) or periodic (alternating). In the framework of linear 
irreversible thermodynamics, we write the constitutive equations
for a polar, active, elastic one-dimensional medium. An analysis of 
the resulting equations for the dynamics of polarity shows that
the transition from graded to alternating polarity patterns is a 
nonequilibrium Lifshitz point. Active contractility is a necessary 
condition for the emergence of sarcomeric, alternating polarity
patterns. 
\end{abstract}

\maketitle

Stress fibers are bundles of actin filaments
made contractile by interaction with myosin minifilaments
and elastic by crosslinking with $\alpha$-actinin 
and other proteins. They are formed in the cytoskeleton of nonmuscle 
animal cells that need either to exert or to resist a mechanical force
(see \cite{Pellegrin2007} for a recent review). 
Their biological function is exemplified \emph{in vivo}
by myofibroblasts that use the contractility of stress fibers to
remodel the extracellular matrix during wound healing
\cite{Tomasek2002}, or by vascular endothelial cells
in blood vessels that align stress fibers parallel to the direction of flow 
to resist shear forces \cite{Drenckhahn1986}. 

Actin filaments are polar structures, with a plus (or barbed) end 
and a minus (or pointed) end. 
The polarity of actin filaments inside a bundle is determined
by decoration of actin with myosin subfragment-1 and 
electron microscopy imaging.
Noncontractile (or passive) actin bundles often present
\emph{uniform} polarity, with barbed ends predominantly facing 
the same direction along the bundle, as seen in filopodia 
\cite{Verkhovsky1997}
or in the retraction fibers of postmitotic cells \cite{Cramer1995}.
The polarity profile of stress fibers
is \emph{alternating}, with a majority
of barbed ends successively pointing in opposite directions
\cite{Drenckhahn1986,Sanger1980,Cramer1997}. 
This periodic structure, with a wavelength of the order of
$1\;\mu$m, is reminiscent of the sarcomeric organization of 
muscle cells. The actomyosin filament bundles
of motile fibroblasts, however,
exhibit a \emph{graded} polarity profile \cite{Cramer1997,Mseka2009}:
The proportion of barbed ends in the bundle facing one direction
varies gradually and monotonically between $0$ and $100 \; \%$ 
from one end to the other. 
The direction of the bundle
sets the orientation of locomotion \cite{Mseka2007}:
The formation of graded polarity bundles is an early
event in the polarization of fibroblasts prior to migration.
The spatial dependence of polarity in contractile actomyosin bundles
contrasts with the uniform polarity observed in structures such as
cilia, filopodia, and retraction fibers,
in which motors do not play an organizing role.

In this Letter, we aim at identifying the physical 
mechanisms responsible for the existence of variable
polarity patterns \cite{Cramer1999} 
in elastic \cite{Deguchi2006,Kumar2006} actin bundles.
We show that \emph{active} contractility, 
the chemomechanical transduction from ATP hydrolysis to motion 
by myosin motors ``pulling'' on actin,
can generate periodic, sarcomerelike polarity patterns 
provided that it exceeds a well defined threshold. 
We focus our attention on \emph{ventral} stress fibers 
\cite{Pellegrin2007} that span the cell ventral surface 
and transmit force to the underlying substrate through end-point 
focal adhesions. We consider existing actomyosin bundles, with constant
total material content, and ignore the intriguing questions raised 
by their assembly \cite{Hotulainen2006}.

\begin{figure}[b]
\includegraphics[width=8cm]{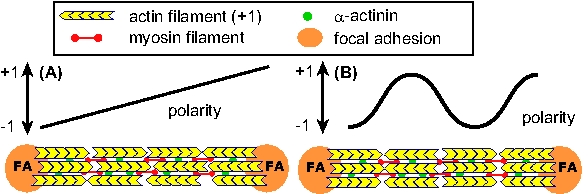}
\caption{\label{fig:model} 
(Color online)
Schematic representation of an actomyosin bundle
with graded (A) and  periodic (B) filament
polarity profile.
}
\end{figure}

Recent theoretical descriptions of stress fibers 
focus on their mechanical properties, \emph{e.g.},
viscoelastic recoil after photoablation 
\cite{Colombelli2009}.
The question of polarity sorting was addressed in
early models of contractile bundles of actin filaments
\cite{Nakazawa1996,*Kruse2003a} that consider a liquid state
in the (dilute) limit of low concentration. 
In a recent extension, density patterns  for viscoelastic  
actomyosin bundles
are studied in the same limit \cite{Peter2008}. 
The effect of coupling the density and polarity fields
in the liquid state is investigated in detail in Ref.~\cite{Kruse2003b}.
Conditions for the spontaneous oscillations of muscle sarcomeres are
determined within the context of active elasticity in Ref.~\cite{Guenther2007}.
We propose to formulate the constitutive equations
for an active, polar, elastic one-dimensional medium, following
the prescriptions of  linear irreversible thermodynamics 
\cite{deGroot1985,Chaikin2000}, valid in principle for
arbitrarily large densities. 
This work is directly inspired by Ref.~\cite{Kruse2005}, where constitutive 
equations were proposed for an active polar viscoelastic liquid
(see \cite{Juelicher2007} for a review).  
Active polar liquid crystals share the same invariance properties 
\cite{Ahmadi2006,Muhuri2007,Giomi2008}. 
We treat here the case of a one-dimensional
viscoelastic solid (Kelvin model), with an 
emphasis on how polar degrees of freedom couple to elastic strain. 
Within this formalism, we restrict our attention 
to lowest-order terms in the free energy and 
in the constitutive equations. 

In the spirit of Ref.~\cite{Kruse2005}, we model a stress fiber as 
a one-component material. The description is mesoscopic,
valid over length scales large compared to the typical mesh size
of the polymer network ($\xi \simeq 10 \; \mathrm{nm}$). 
The geometry is one-dimensional: The stress
fiber of (constant) length $L$ occupies the interval $x \in [-L/2, \; L/2]$.  
Our hydrodynamic description is built on
three conserved fields: the mass, momentum and energy densities 
$\rho(x,t)$, $\vec{g}(x,t)$, and  $u(x,t)$, respectively; and one slow mode,
the polarity $\vec{p} = p(x,t) \; \vec{e}_x$, defined as the coarse-grained 
average of the polarity of actin filaments present at time $t$ in the 
vicinity of $x$. By convention, $p = +1$ (respectively, $p = -1$) corresponds  to
a bundle where all (respectively, no) barbed ends point towards positive $x$
(see Fig.~\ref{fig:model}). 
The relevant boundary conditions for the polarity field are
``antisymmetric'': 
\begin{equation}
  \label{eq:BC}
p(-L/2) = -1, \;\;p(L/2) = +1  
\end{equation}
since the barbed ends of actin filaments face the end-point 
focal adhesions \cite{Verkhovsky1997,Cramer1997}.

The stress fiber is a viscoelastic solid \cite{Kumar2006}, 
with velocity and strain fields $v(x,t)$  and $e(x,t)$.
We focus on time scales for polarity ordering that are long compared 
to the typical turnover time of actin and other dynamic protein components:
$t \gg \tau_{\mathrm{TO}} \simeq 10^2 \; \mathrm{s}$
\cite{Hotulainen2006}. 
The density field $\rho$ is a fast variable, slaved to the strain field:
$\delta \rho / \rho = - e$.
The free energy density, expanded 
to quadratic order about $p = e = 0$,  reads
\footnote{Since the boundary conditions on $p$ are fixed, we may
ignore the polar term $\partial_x p$ in the free energy density.}:
\begin{equation}
  \label{eq:freeenergy}
  f = \frac{a}{2} \; p^2 + 
      \frac{K}{2} \; \left( \partial_x p \right)^2 +
      \frac{G}{2} \; e^2+
      w \; e \; \partial_x p,
\end{equation}
where all terms are invariant under inversion of space and polarity. 
The positive coefficients $a$, $K$, and $G$
respectively denote the susceptibility, 
the energy cost of inhomogeneities of the polarity field,
and the elastic modulus. 
Symmetries of the problem 
allow for a coupling term between strain and thepolarity gradient,
with coupling parameter $w$ of unknown sign. 
From Eq.~(\ref{eq:freeenergy}), we obtain the molecular field $h$,
conjugate to the polarity: 
$h = -\frac{\delta f}{\delta p} = - \; a \; p 
+ K \; \partial^2_x p +w \; \partial_x e$
and the elastic stress 
$\sigma^{\mathrm{el}} = \frac{\delta f}{\delta e} = G \; e + w \; \partial_x p$.
Thermodynamic stability imposes the constraint $w^2 \le K \; G$:
The uniform, strain-free, zero polarity
steady-state is a stable solution when boundary conditions permit. 

We proceed to derive the dissipation rate $R$:
\begin{equation}
  \label{eq:R}
  R = \left( \sigma + P - \sigma^{\mathrm{el}} \right) \; \partial_x v 
      + \dot{p} \; h + r \; \Delta \mu,
\end{equation}
where $P$ and $\sigma$ denote the pressure and (total) stress fields, 
respectively.
Equation (\ref{eq:R}) contains a mechanical term, a polar term 
($\dot{p}$ is the total derivative of $p$), and an active term where 
the chemical force $\Delta \mu$ is conjugate to the ATP 
consumption rate $r$ \cite{Kruse2005}.   
We next express the relevant generalized thermodynamic fluxes as linear
functions of the generalized thermodynamic forces and obtain
the stress fiber's constitutive equations:
\begin{eqnarray}
  \label{eq:consteqp}
  \dot{p} &=& \Gamma \; h - \alpha  \; p \partial_x p \; \Delta \mu\\
  \label{eq:consteqsig}
  \sigma  +  P - \sigma^{\mathrm{el}} &=&  \eta \; \partial_x v
           + (- \zeta + \beta \; \partial_x p) \; \Delta \mu
\end{eqnarray}
The positive kinetic coefficient $\Gamma$ is analogous to 
an inverse viscosity \cite{deGennes1995}, and 
$\eta$ is the viscosity of the fiber.
A polar, nonrelaxational term with a coupling coefficient $\alpha$
of unknown sign is included in Eq.~(\ref{eq:consteqp}), in agreement with
studies of active liquid crystals 
\cite{Giomi2008,Ahmadi2006,Muhuri2007}.
The last term in Eq.~(\ref{eq:consteqsig}), a nondiagonal
chemomechanical coupling, gives the \emph{active}
contribution to the total stress. The coefficient $\zeta$ is
negative for contractile actomyosin \cite{Kruse2005}. 
The invariance properties
of polar media allow for an active coupling between 
stress and polarity gradient, with a coefficient
$\beta$, as first introduced in Ref.~\cite{Giomi2008}.
For simplicity, we ignore the punctate spatial pattern of myosins
in stress fibers \cite{Pellegrin2007,Cramer1997}: $\Delta \mu$ is a
constant in the following.

We further assume that the time scale for polarity ordering is much 
longer than that for strain relaxation, of the order of $1$ s as measured in 
laser ablation experiments \cite{Kumar2006}. 
In this long time limit the velocity field $v$ 
and its gradient are negligible.
The pressure field is uniform and equal to the pressure in the
surrounding cytosol $P = P_0$. 
Newton's law, $\partial_x \sigma = 0$, yields:
$G \; \partial_x e + \left( w + \beta \; \Delta \mu \right) \partial^2_x p = 0$,
which eliminates the strain field $e$ in Eq.~(\ref{eq:consteqp}). 
This yields the evolution equation for the polarity field.

In the \emph{passive} case $\Delta \mu = 0$, 
the total derivative $\dot{p}$ reduces to $\partial_t p$ 
when $v = 0$. We obtain a
damped diffusion equation for the polarity field:
\begin{equation}
  \label{eq:passive}
  \partial_t p = - a \Gamma \; \left( p - l^2 \; \partial^2_x p \right),
\end{equation}
where the length scale  
$l^2 = (K/a) \left[1 - w^2/(K G) \right] \ge 0$
has been introduced. 
Using antisymmetric boundary conditions,
we find a linearly stable,
stationary, monotonic solution of Eq.~(\ref{eq:passive}):
$p(x) =  \sinh\left(x/l\right)  / \sinh\left(L/2 l\right)$
that describes the graded polarity profile of 
actin bundles in the absence of myosin-induced contractility.

\begin{figure}[t]
\includegraphics[width=8cm]{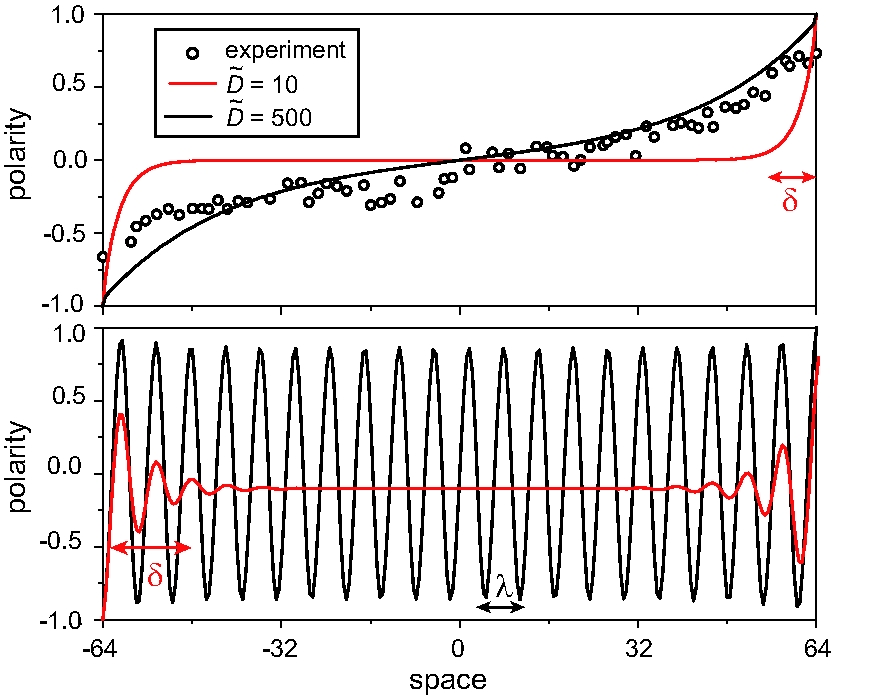}
\caption{\label{fig:profile} 
(Color online)
Stationary polarity profiles obtained from  
numerical simulations. 
We set the dimensionless system size to $\tilde{L} = 128$ and use the
boundary conditions (\ref{eq:BC}).
Top: When $\tilde{\nu} = 0$ and $\tilde{D} = 500$,
the graded profile obtained 
for the damped Burgers equation (black line)  
agrees qualitatively with experimental data, taken from Cramer et al. 
\cite{Cramer1997},
where the first and last data points are set at the boundary (circles).  
Mixed polarity is obtained far from the tips when $\tilde{D} = 10$
(red line).
Bottom: Typical periodic 
profile for the damped KS equation with $\tilde{\nu} = 1.0$ and
$\tilde{D} = - 2.01$ (black line). 
Mixed polarity is obtained far from the tips with $\tilde{\nu} = 1.0$ and 
$\tilde{D} = - 1.90$ (red line).
}
\end{figure}

In the \emph{active} case $\Delta \mu > 0$, 
Eq.~(\ref{eq:consteqp}) yields a damped Burgers equation:
\begin{equation}
  \label{eq:activegraded}
  \partial_t p + \alpha  \Delta \mu \; p \partial_x p = - a \Gamma p
+ D  \; \partial^2_x p ,
\end{equation}
with a diffusion constant 
$D = \Gamma K \; \left[ 1 - 
   \frac{w^2}{K G} \left( 1 + \frac{\beta \Delta \mu}{w} \right)  \right]$.
For \emph{weak} activity, \emph{i.e.} when 
$\beta \Delta \mu/w < K G/w^2 - 1$,
$D$ is positive, and 
monotonic stationary solutions qualitatively similar to 
the above analytical expression are obtained 
numerically (see Fig.~\ref{fig:profile}). 
These linearly stable solutions describe the
graded polarity patterns observed in contractile
stress fibers.

For \emph{strong} activity, \emph{i.e.} when 
$\beta \Delta \mu/w > K G/w^2 - 1$, the diffusion coefficient
changes sign to become negative. In the particular case $\alpha = 0$, we obtain
periodic stationary solutions:
$  p(x) = \sin\left(q x\right)/  \sin\left(q L / 2\right)  $,
with a wave number $q^2 = - a \Gamma/D$.
Since $D <0$, stationary solutions of Eq.~(\ref{eq:activegraded}) 
are linearly unstable. 
We are led to include a stabilizing, higher-order 
polarity gradient term $\nu/2 \; \left( \partial^2_x p \right)^2$ 
in the free energy  density, where $\nu$ is a positive higher-order 
diffusion constant. This is reminiscent of a Lifshitz point
\cite{Chaikin2000,deGennes1995}, where an equilibrium phase transition 
between a  uniform and a modulated phase occurs as a diffusion coefficient 
changes sign. The field $h$ is supplemented with
the term $- \nu \;  \partial^4_x p$ 
and Eq.~(\ref{eq:consteqp}) becomes a damped Kuramoto-Sivashinsky (KS) equation:
\begin{equation}
  \label{eq:dampedKS}
  \partial_t p + \alpha  \Delta \mu \; p \partial_x p = - a \Gamma p
+ D \; \partial^2_x p 
- \nu \Gamma \; \partial^4_x p,
\end{equation}
For large enough damping (here $a \Gamma$), this equation 
possesses stable stationary solutions periodic in space \cite{Manneville1990}.
The transformation
$\tilde{t} = a \Gamma t$, $\tilde{x} = a \Gamma x / (|\alpha| \; \Delta \mu)$,
makes Eq.~(\ref{eq:dampedKS}) dimensionless, with only two
parameters 
$\tilde{D} = \frac{a \Gamma D}{(\alpha \Delta \mu)^2}$
and
$\tilde{\nu} = \frac{\nu}{a} (\frac{ a \Gamma}{\alpha  \Delta \mu })^4$.
Alternating polarity patterns are obtained when the uniform solution
$p(x) = 0$ is linearly unstable, \emph{i.e.}
for $\tilde{D} < \tilde{D}_c = - 2\sqrt{\tilde{\nu}}$
(see Fig.~\ref{fig:profile}). Their wavelength $\lambda$ is obtained 
where the growth rate of the linearized damped KS equation is maximal:
\begin{equation}
 \label{eq:wavelength}
\lambda^2 =  - \frac{8 \pi^2 \Gamma \nu}{D} = \frac{8 \pi^2 \nu}{K} 
\left[ \frac{w^2}{K G} \left( 1 + \frac{\beta \Delta \mu}{w} \right) -1
\right]^{-1}.
\end{equation}
We predict that the inverse square wavelength $\lambda^{-2}$
of alternating polarity patterns
is a linear function of $\Delta \mu$, as may be checked 
experimentally by up- or down-regulating the 
contractility of fibers. The wavelength increases with $G$ and is
independent of  $\alpha$, and also of $L$ when $L \gg \lambda$.  
We checked numerically that the periodicity of patterns is robust 
to sufficiently weak spatial modulations of $\Delta \mu$ 
mimicking the punctate profile of myosins, 
as well as to the addition of a noise term of small amplitude.

\begin{figure}[t]
\includegraphics[height=5cm]{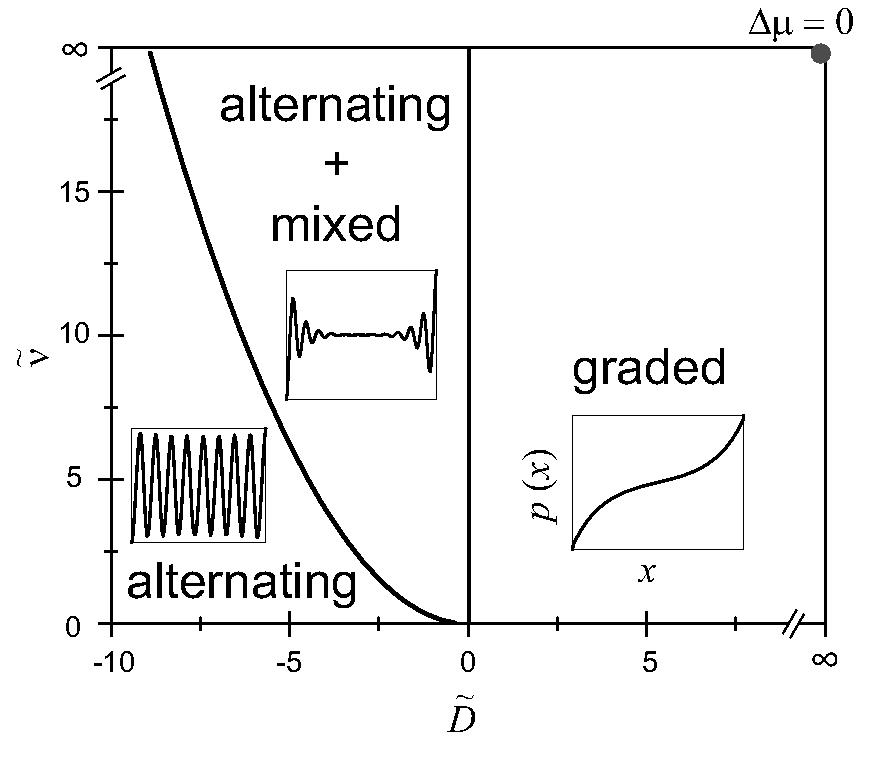}
\caption{\label{fig:phasediag} 
Phase diagram of an active fiber in the $(\tilde{D}, \tilde{\nu}$) plane,
with $\tilde{D} = \frac{a \Gamma D}{(\alpha \Delta \mu)^2}$ and
$\tilde{\nu} = \frac{\nu}{a} (\frac{ a \Gamma}{\alpha  \Delta \mu })^4$.
The polarity profile is graded in the passive case 
($\tilde{D}$ and $\tilde{\nu} \rightarrow + \infty$ when 
$\Delta \mu \rightarrow 0$).
}
\end{figure}

The case of ``mixed'' polarity  ($p = 0$) is also 
accounted for \cite{Cramer1999}. 
When $D>0$, the uniform solution  $p(x) = e(x) = 0$ is stable but irrelevant 
because of incompatible boundary conditions. Mixed polarity may, however, be 
observed locally, far from the tips, when the penetration
length $\delta \propto \sqrt{D}$ 
is small compared to the total length $L$ (see Fig.~\ref{fig:profile}).
When $D<0$,  mixed polarity is also obtained far from the tips 
in the range $\tilde{D}_c < \tilde{D} < 0$, 
with \emph{alternating} polarity near the boundaries.
Simulations suggest that the penetration length
is of the order of $\delta \propto (\tilde{D} - \tilde{D}_c)^{-1/2}$. 
To the best of our knowledge, this type of pattern 
has not been observed in experiments.
These results are summarized in a phase diagram (Fig.~\ref{fig:phasediag}).

In the above derivation, we implicitly assumed that polarity 
is a \emph{nonconserved} field. 
By disregarding the possibility of spontaneous polarity reversal
of a filament, which we deem extremely unlikely in a 
one-dimensional cross-linked bundle, this implies
that relevant time scales are longer than the characteristic time
$\tau_{\mathrm{C}}$ for nucleation/annihilation and insertion/removal
of actin filaments. Equations (\ref{eq:passive}) and (\ref{eq:dampedKS}) 
are valid only when $t > \tau_{\mathrm{C}}$.

We now derive the constitutive equations for a \emph{conserved} polarity field,
valid for shorter time scales $t < \tau_{\mathrm{C}}$
(see \cite{Kruse2003b} for an alternative approach).
The conservation equation reads $\dot{p} =  - \partial_x j^p$, and $j^p$, 
the polarity current, is the flux conjugate to the force $\partial_x h$.
The constitutive equations read:
\begin{eqnarray}
j^p &=&   \Gamma \; \partial_x h
+ (-\bar{\zeta} + \bar{\beta} \; \partial_x p + \alpha \; p^2) \; \Delta \mu   
- \chi \; \partial_x v \nonumber\\
  \sigma  &=&  - P + \sigma^{\mathrm{el}} + \eta \partial_x v
           + (- \zeta + \beta \partial_x p) \Delta \mu
           + \chi \partial_x h \nonumber
\end{eqnarray}
where $\bar{\zeta}$, $\bar{\beta}$, and $\chi$ are additional
cross-coupling coefficients of unknown sign and 
the nonlinear term  $\alpha  p^2  \Delta \mu$ is included
in analogy with the nonconserved case.
An Onsager relation sets the reactive nondiagonal coupling between stress and
the gradient of the molecular field.

For a \emph{passive} bundle ($\Delta \mu = 0$), we obtain
the differential equation obeyed by stationary polarity profiles:
$\partial^2_x \left( p - l^2 \; \partial^2_x p \right) = 0$,
in the limit where $v = \partial_x v = 0$.
Linear stability of the uniform, zero polarity steady-state 
imposes the condition $w \chi > 0$.
For antisymmetric boundary conditions [Eq.~(\ref{eq:BC})],
we find as a linearly stable
stationary solution the linear profile
$p(x) = 2 x/L$.

For an \emph{active} bundle ($\Delta \mu > 0$),
we obtain a set of two coupled equations for $p$ and $h$: 
\begin{eqnarray}
  \label{eq:cons1}
 \partial_t p + \alpha  \Delta \mu \; p \; \partial_x p
 &=& - \Gamma \; \partial^2_x h  
- \bar{\beta} \Delta \mu \; \partial^2_x p,
\\  
\label{eq:cons2}
 \left( 1 - \frac{w \chi}{G} \; \partial^2_x \right) h 
 &=& - a p + \frac{D}{\Gamma}  \; \partial^2_x p 
 - \nu \; \partial^4_x p,
\end{eqnarray}
where the higher-order term  $\nu/2 \; \left( \partial^2_x p \right)^2$ 
was included in the free energy density. 
An exhaustive study of the phase diagram of the system 
(\ref{eq:cons1}-\ref{eq:cons2}) is beyond 
the scope of this Letter. We naturally expect the graded polarity patterns
obtained in the passive case to be preserved for weak enough activity.
We shall restrict our analysis to the limit
$w \chi >> G$, where the molecular field may be eliminated.
In this limit, we obtain a damped KS equation:
\begin{equation}
  \label{eq:dampedKScons}
  \partial_t p + \alpha  \Delta \mu \; p \partial_x p = 
- \epsilon_{\mathrm{c}} \;  p + D_{\mathrm{c}} \; \partial^2_x p 
- \nu_{\mathrm{c}} \; \partial^4_x p
\end{equation}
with a damping coefficient $\epsilon_{\mathrm{c}} = G  \Gamma a/w \chi$,
a diffusion constant $D_{\mathrm{c}}$ given by: 
$D_{\mathrm{c}} = \frac{\Gamma G K}{w \chi} \left(1 - \frac{w^2}{K G}  \right)
- \Delta \mu \; \left(\frac{\Gamma \beta}{\chi} + \bar{\beta}\right)$,
and a higher-order diffusion constant
$\nu_{\mathrm{c}} =  G  \Gamma \nu / w \chi$.
In this regime, beyond the nonequilibrium Lifshitz point 
($D_{\mathrm{c}} < 0$), and
for large enough damping  $\epsilon_{\mathrm{c}}$,
we shall again find alternating polarity patterns.
The transition may be driven either by the active stress
(coefficient $\beta$) or by the active current
(coefficient $\bar{\beta}$).

A key ingredient of our model of an active, polar, one-dimensional 
elastomer is the coupling between the elastic strain
and polarity gradient, allowed by symmetry in the free energy expansion
[Eq.~(\ref{eq:freeenergy})].
We show that an alternating polarity profile emerges in an active medium
with a \emph{ uniform} spatial distribution of myosins: 
Coupling with the myosin density field is expected to
lead to myosin patterns with the same wavelength.
While graded polarity patterns are allowed for both 
passive and active bundles, active contractility is a necessary condition
for the self-organization of stress fibers into 
sarcomerelike, 
alternating polarity patterns.
The alternating polarity pattern of
an active actomyosin bundle may therefore become graded 
upon treatment with drugs that inhibit contractility. 
Whereas alternating profiles occur far from the boundaries
beyond the instability threshold, graded profiles 
are made possible by nonzero values of the polarity field at the tips.
These conclusions hold irrespective of the conserved or  nonconserved nature
of the polarity field. 

Linear irreversible thermodynamics is a 
power\-ful fra\-me\-work within which
coupling terms are deemed re\-le\-vant according to their 
invariance properties. 
The numerical values of coupling constants are beyond
the scope of the theory and may well depend here on the cell type. 
The only measurement we are aware of is that of
the elastic modulus $G \simeq 10^5-10^6$ Pa \cite{Deguchi2006}. 
Giving a microscopic interpretation to hydrodynamic coefficients such as 
$w$, which couples elastic and polar degrees of freedom, will require
models that relate behavior at hydrodynamic scales
with the microscopic  interaction of actin filaments 
with active and passive cross-linkers.
One possible interpretation of $w$ is that a 
preference of cross-linkers to parallel or
antiparallel pairs of actin filaments
may lead to a different value of the stress at constant strain.

\bibliography{polarity}

\end{document}